\documentclass[aps,prd,showpacs]{revtex4}
\usepackage[T1]{fontenc}
\usepackage[latin9]{inputenc}
\usepackage{color}
\usepackage{amsmath}
\usepackage{graphicx}
\usepackage{amssymb}

\newcommand{\CD}{{\cal D}}
\newcommand{\CE}{{\cal E}}
\newcommand{\CF}{{\cal F}}

\newcommand{\CQ}{{\cal Q}}
\newcommand{\CR}{{\cal R}}

\newcommand{\CW}{{\cal W}}
\newcommand{\CM}{{\cal M}}
\newcommand{\average}[1]{\left\langle #1 \right\rangle_\CD}

\newcommand{\initial}[1]{{#1_{\rm \bf i}}}

\begin{document}
\title[Towards physical cosmology]{Towards physical cosmology:\\ geometrical interpretation of Dark Energy, Dark Matter and Inflation\\ without fundamental sources\footnote{presented at the International Conference on Two Cosmological Models, Universidad Iberoamericana Ciudad de M\'exico -- Department of Physics and Mathematics, November 19, 2010.}}
\author{Thomas Buchert}
\affiliation{Universit\'e Lyon~1, Centre de Recherche Astrophysique de Lyon,\\
CNRS UMR 5574, 9 Avenue Charles Andr\'e, F--69230 Saint--Genis--Laval,
France\\
Email: buchert@obs.univ--lyon.fr\\}
%
\smallskip
%

\bigskip\bigskip

\begin{abstract}
We outline the key--steps towards the construction of a physical, fully relativistic cosmology, in which we aim to trace Dark Energy and Dark Matter back to physical properties of space. The influence of inhomogeneities on the effective evolution history of the Universe is encoded in backreaction terms and expressed through spatially averaged geometrical invariants. These are absent and interpreted as missing dark fundamental sources in the standard model. In the inhomogeneous case they can be interpreted as energies of an emerging scalar field (the morphon). These averaged invariants vanish for a  homogeneous geometry, where the morphon is in an unstable equilibrium state. If this state is perturbed, the morphon can act as a classical inflaton in the Early Universe and its de--balanced energies can mimic the dark sources in the Late Universe, depending on spatial scale as Dark Energy or as Dark Matter, respectively.
We lay down a line of arguments that is qualitatively conclusive, and we outline open problems of quantitative nature, related to the interpretation of observations.
\end{abstract}
\maketitle

\tableofcontents

\clearpage

\section{General relativity and cosmology}

\subsection{The foliation issue and the notion of an effective cosmology}

The homogeneous--isotropic standard model of cosmology, being itself a particular solution of Einstein's general theory of relativity, does by far not exploit the degrees of freedom inherent in the geometry as a dynamical variable. It is this richer tone of general relativity -- as compared to the Newtonian theory -- that opens the possibility to generalize cosmological models, notably by including inhomogeneous structure also in the geometrical variables.  There are several guidelines to be emphasized in such a generalization: firstly, a cosmology is thought of as an evolving space section that implies the need to speak of a foliated space time, introducing four degrees of freedom (the lapse and shift functions in an ADM setting). This necessarily implies, on general grounds, a breaking of four--dimensional covariance. This fact should not be confused with coordinate-- or gauge--dependence of the resulting cosmological equations and variables, however. Secondly, a cosmology 
purports an effective point of view in the sense that the evolving spatially inhomogeneous variables are thought of as being ``averaged over'' in a way that has to be specified.
We aim at a description that only implicitly refers to a metric. However, if a metric is to be specified,
a cosmological metric is then to be considered as an effective, ``smoothed out'' or {\it template metric}, being not necessarily a solution of the equations of general relativity. 
Finally, a {\it physical} cosmology should be characterized by such an effective evolution model, an effective metric to provide the distance scale for the interpretation of  observations, or alternatively an evolution model for average characteristics on the light cone, together with a set of initial data. These latter are to be related to physical properties of fundamental sources, but also to the geometrical data at some initial time (effective, i.e ``averaged'' quantities of known energy sources, intrinsic and extrinsic curvature). This latter clearly emphasizes the absence of any phenomenological parameters. Those would just parametrize our physical ignorance. All these points will be made explicit in what follows.

\subsection{The dark side of the standard model: postulated sources and proposed solutions}

The high level of idealization of the geometrical properties of space in the standard model leads to the need of postulating sources that would generate ``on average'' a strictly, i.e. globally and locally, homogeneous geometry. It is here where a considerable price has to be paid for a model geometry that obviously is not enough to meet physical reality, unless we really believe that we can find the missing sources: $96$ percent of the energy content is missing in the form of a) a postulated source acting attractive like matter, so--called Dark Matter ($\cong 23$ percent) and b) a postulated source acting repulsive, so--called Dark Energy ($\cong 73$ percent). Evidence for the former does indeed come from various scales (galaxy halos, clusters and cosmological, see e.g. \cite{roos}), while evidence for the latter only comes from the apparent magnitude of distant supernovae (see \cite{SNIa:Union,SNIa:Constitution,SNIa:Essence} for the latest data) that, if interpreted within standard model distances, would need an accelerating model. In the simplest case this volume acceration is achieved by a homogeneous--isotropic cosmology with a cosmological constant. 
It should be emphasized that when we speak of evidence, we already approach this evidence with model priors \cite{huntsarkar,seikel:acc,cmbobs}.
Keeping this idealization for the geometry of the cosmological model for example, one has to conjecture fundamental fields in proportion to the missing dark components on cosmological scales. The search for these fields is one major research direction in modern cosmology. 

\smallskip

Another huge effort is directed towards a generalization of the underlying theory of gravitation. While this would generalize the geometry of the model, it is not clear why all these efforts go into a generalization of general relativity and not into the generalization of the cosmological model within general relativity. There are certainly good lines of arguments and various motivations in particle physics and quantum gravity to go beyond the theory of Einstein (for reviews see \cite{DE:review}, \cite{DE:pilar}), but the ``dark problem'' may be first a classical one. 

\smallskip

Looking at generalizations of the standard model \emph{within} general relativity can be identified as a third research direction to which we dedicate our attention here. In light of current efforts it is to be considered conservative, since it does not postulate new fundamental fields and it does not abandon a well--tested theory of gravitation \cite{dressing}, \cite{rasanen:de}, \cite{kolb:backreaction} (for reviews on the physical basis of this third approach see \cite{buchert:jgrg,buchert:review} and \cite{rasanen:acceleration}). 
Among the works in this latter field, research that analyzes spherically symmetric exact solutions has been meanwhile developed to some depth, and has determined the constraints, necessary to explain Dark Energy, on a postulated observer's position within a large--scale void (see \cite{LTB:review,bolejkoandersson,celerier,voidtest}
and references therein).  

\subsection{Fictitious and physical backgrounds}

Perhaps a reason for not questioning the standard model geometry within general relativity and to go for the search for fundamental fields or for generalizations of the laws of gravitation is the following belief: effectively, i.e. ``on average'', the model geometry has to be {\it homogeneous}, since structures should be ``averaged over''. Then, due to observational facts on large scales (the high degree of isotropy of the Cosmic Microwave Background, if the dipole is completely eliminated due to our proper motion with respect to an idealized exactly isotropic light sphere), and first principle priors (the {\it strong cosmological principle} that requires the universe model to look the same in all directions), the model geometry is taken to be {\it locally isotropic}. 

\smallskip

Taking this reasoning at face value we must note two points: the notions of homogeneity and isotropy in the standard model are both too strong to be realistic: firstly, local isotropy implies a model that is locally and globally homogeneous, i.e. despite the conjecture that the homogeneous model describes the inhomogeneous Universe ``on average'', this {\it strict homogeneity} does not account for the fact that any averaging procedure, in one way or another, would introduce a {\it scale--dependence} of the averaged (homogeneous) variables \cite{ellisbuchert}. This scale--dependence, inherent in any physical averages, is suppressed. Even if a large {\it scale of homogeneity} exists (we may call this {\it weak homogeneity principle}), the model is in general scale--dependent on scales below this homogeneity scale \cite{sylos:copernican}. The same is true for isotropy: while the averaged model may be highly isotropic on large scales, a realistic distribution on smaller scales is certainly not (we may call this {\it weak isotropy principle}). Correspondingly, a {\it weak cosmological principle} would be enough to cover the reality needs while still facing observational evidence on large scales.  

\smallskip

We may summarize the above thoughts by noting that, on large scales, a homogeneous--(almost)isotropic {\it state} does not necessarily correspond to a homogeneous--(almost)isotropic {\it solution} of Einstein's equations. These former states are the averages over fluctuating fields and it is only to be expected that the state coincides with a strictly homogeneous solution in the case of absence of fluctuations. In other words, looking at fluctuations first requires to establish the average distribution. Only then the notion of a {\it background} makes physical sense \cite{kolb:backgrounds}. Current cosmological structure formation models, perturbation theories or N--body simulations, are constructed such that the average vanishes on the background of a homogeneous--isotropic {\it solution} \cite{buchertehlers}.  A such chosen reference background may be a {\it fictitious background}, since it arises by construction rather than derivation. On the contrary, a {\it physical background} is one that corresponds to the average (whose technical implementation has to be specified, and which is nontrivial if tensorial quantities like the geometry have to be ``averaged''). A sound implementation of a physical background will be a statistical background where not only solutions but ensembles of solutions are averaged.
Having specified such an averaging procedure, a physical cosmological model may then be defined as an evolution model for the average distribution. Despite these remarks it is of course possible that the homogeneous solution forms at the same time the average. A well--known example is Newtonian cosmology \cite{buchertehlers}. It is also conceivable that the homogeneous solution provides, in some spatial and temporal regimes, a good approximation for the average. Still, it is important to consider perturbations on the correct background solution \cite{kolb:voids}.

\section{Scalar field models and the morphon}

\subsection{Effective evolution of inhomogeneous universe models}

Taking the point of view of generalizing the cosmological model within general relativity by abandoning the strong cosmological principle (strict homogeneity and isotropy on all scales) and replacing it by the weak cosmological principle (existence of a homogeneity scale and restriction to effective states that are almost isotropic on the scale of homogeneity) leads us to a ``rewriting of the rules'' to build the cosmological model. We shall consider the rules that led to the standard model of cosmology and replace them by their more general counterparts. It follows a basically similar framework that displays, however, a signature of inhomogeneity through the occurence of so--called backreaction terms and through a manifest scale--dependency. We shall not introduce new principles or assumptions, apart from the above outlined relaxation of the cosmological principle. We shall restrict ourselves to the simplest case of an irrotational dust model (for generalizations of the dust model \cite{buchert:dust} with non--constant lapse function see \cite{buchert:fluid}, and for additionally non--vanishing shift see \cite{brown1,brown2,larena,veneziano2}).

\medskip\noindent
$\bullet$ As in the standard model we introduce a foliation of space time into flow--orthogonal hypersurfaces. We generalize the notion of {\it Fundamental Observers} to those that are in free fall also in the general space time. Although, as in the standard model, this setup depends on the chosen foliation, we presume that this choice is unique as it prefers the fundamental observers against observers that may be accelerated with respect to the hypersurfaces. A general inhomogeneous hypersurface -- contrary to the homogeneous case -- will, in this setting, unavoidably run into singularities in the course of evolution. This is to be expected in a given range of spatial and temporal scales, since we are treating the matter model as {\it dust}.
This is not a problem of the chosen foliation, but a problem of the matter model that has to be generalized, if small--scale structure formation has to remain regular, and this can be achieved by the inclusion of velocity dispersion and vorticity. 

\medskip\noindent
$\bullet$ As in the standard model we confine ourselves to scalar quantities. We replace, however, the homogeneous quantities by their spatial averages, e.g. the homogeneous density $\varrho_H (t)$ is replaced by $\average{\varrho} (t)$ for the inhomogeneous density $\varrho$ that is volume--averaged over some compact domain $\CD$.
We realize the averaging operation by a mass--preserving Riemannian volume average. In some mathematical disciplines and in statistical averages at one instant of time, it may be more convenient to introduce a volume--preserving averager, but thinking of an averaging domain that is as large as the homogeneity scale we have to preserve mass rather than volume. Furthermore, the average is performed with respect to the above--defined {\it Fundamental Observers}. Spatially averaging a scalar $\Psi (t,X^i)$, as a function of Gaussian coordinates $X^i$ and a synchronizing time $t$, is defined as: 
\begin{equation}
\label{average}
\langle \Psi (t, X^i)\rangle_{\cal D}: = 
\frac{1}{V_{\cal D}}\int_{\cal D}  \;\Psi (t, X^i) \;d\mu_g\;\;\;,
\end{equation}
with the Riemannian volume element $d\mu_g := \sqrt{g} d^3 X$, $g:=\det(g_{ij})$, and 
the volume of an arbitrary compact domain, $V_{\cal D}(t) : = \int_{\cal D} \sqrt{g} d^3 X$.
Note that within a more general setup that includes lapse and shift functions, we would have to consider the question whether the locally lapsed time is replaced by a global ``averaged time'' that would involve an average over the lapse function. Here, the dust cosmology is already synchronous, so that this question does not arise. 
Note furthermore, that the building of averages is done in the inhomogeneous geometry. The averages functionally depend on the inhomogeneous metric, but this latter needs not to be specified. We may talk of a kinematical averaging that does not deform the geometry, i.e. that does not change the physical properties of the inhomogeneous space time.
For other strategies, see \cite{ellisbuchert}, and references therein, as well as Section~IV.

\medskip\noindent
$\bullet$ We generalize the kinematical laws of the standard model a) for the volume expansion (the Hamiltonian constraint in the ADM formulation of general relativity) and b) for the volume acceleration (Raychaudhuri's equation in the ADM formulation of general relativity) by dropping the symmetry assumption of local isotropy. The general equations are then volume--averaged,
leading to the following general volume expansion and volume acceleration laws (for a volume scale factor, defined by $a_{\CD}\left(t\right):=\left(V_{\cal D}(t) / V_{\cal D}(t_{i})\right)^{1/3}$; the overdot denotes partial time--derivative, which is the covariant time--derivative here) \cite{buchert:dust}:
\begin{equation}
3\frac{\ddot{a}_{\CD}}{a_{\CD}}  =  -4\pi G\average{\varrho}+\CQ_{\CD}+\Lambda\;\;\;\;\;;\;\;\;\;\;
3H_{\CD}^{2} + \frac{3 k_\CD}{a_\CD^2} = 8\pi G\average{\varrho} -\frac{1}{2}\CW_\CD - \frac{1}{2}\CQ_{\CD}+\Lambda\;,
\label{averagedequations}
\end{equation}
where $H_{\CD}$ denotes the domain dependent Hubble rate $H_{\CD}=\dot{a}_{\CD} / a_{\CD}=-1/3\average{K}$, $K$ is the trace of the extrinsic curvature of the embedding of the hypersurfaces into the space time, $K_{ij}$, and $\Lambda$ the cosmological constant.
The {\it kinematical backreaction} $\CQ_{\CD}$ is composed of averaged extrinsic curvature invariants, while $\CW_\CD$ is an averaged intrinsic curvature invariant 
that describes the deviation of the average of the full (three--dimensional) Ricci scalar curvature $\CR$ from a constant--curvature model,
\begin{equation}
\CQ_{\CD}:= \average{K^2 - K^i_{\;j}K^j_{\;i}} - \frac{2}{3}\average{K}^2 \;\;\;\;;\;\;\;\;\CW_{\CD}: = \average{\CR} - \frac{6 k_\CD}{a_\CD^2}\;.
\label{eq:Def-Q}
\end{equation}
The kinematical backreaction $\CQ_{\CD}$ can also be expressed in terms of kinematical invariants, where the extrinsic curvature is interpreted actively in terms of (minus) the expansion tensor:
\begin{equation}
\CQ_{\CD}:=\frac{2}{3}\left(\average{\theta^{2}}-\average{\theta}^{2}\right)-2\average{\sigma^{2}}\;,
\label{eq:Def-Q}
\end{equation}
where $\theta$ is the local expansion rate and $\sigma^{2}:=1/2 \sigma_{ij}\sigma^{ij}$
is the squared rate of shear. Note that $H_{\CD}$ is now defined as $H_{\CD}=1/3\average{\theta}$.
$\CQ_{\CD}$ appears as a competition term between the 
averaged variance of the local expansion rates, $\average{\theta^{2}}-\average{\theta}^{2}$,
and the averaged square of the shear scalar $\average{\sigma^{2}}$ on the domain
under consideration. 

\smallskip

For a homogeneous domain the above backreaction terms $\CQ_\CD$ and $\CW_\CD$, being covariantly defined and gauge invariants in a perturbation theory on a homogeneous background solution, are zero. They encode the departure from homogeneity in a coordinate--independent way \cite{gaugeinv,veneziano2}.

\smallskip

The integrability conditions connecting the two Eqs.~(\ref{averagedequations}), assuring that the expansion law is the integral of the acceleration law, read:
\begin{equation}
\langle\varrho{\dot\rangle} + 3 H_\CD \average{\varrho} \;=0\;\;\;;\;\;\;\;
a_{\CD}^{-2}( a_{\CD}^{2}{\CW}_\CD {\dot )}\;+\;a_{\CD}^{-6}( a_{\CD}^{6}{\CQ}_{\CD} {\dot )}\;=0\;.
\label{eq:integrability}
\end{equation}
While the mass conservation law for the dust is sufficient in the homogeneous case, there is a further equation connecting averaged intrinsic and extrinsic curvature invariants in the inhomogeneous case. The expressions in brackets are conformal invariants (for further details see \cite{buchert:review}.

\smallskip

The interpretation of these average equations as {\it generalized or evolving backgrounds} \cite{buchert:review}, \cite{kolb:backgrounds} implies that the second conservation law describes an interaction between structure formation and background curvature. In the standard model this latter is absent and structures evolve independently of the background. This constant--curvature background furnishes the only solution of (\ref{eq:integrability}), in which structure formation decouples from the background (the expressions in brackets in the second conservation law are separately constant).
Backreaction on such a fixed background decays in proportion to the square of the density and is unimportant in the Late Universe \cite{buchert:dust, buchert:darkenergy, buchert:review}. This  degenerate case of a decoupled evolution explains the fact that in Newtonian and quasi--Newtonian models backreaction has no or little relevance \cite{buchert:review}; in the Newtonian case \cite{buchertehlers}, as well as in Newtonian \cite{bks,abundance} and spatially flat, relativistic spherically symmetric dust solutions \cite{singh1}, $\CQ_\CD$ vanishes. In models with homogeneous geometry and with periodic boundary conditions imposed on the inhomogeneities on some scale, the backreaction term is globally zero and describes cosmic variance of the kinematical properties.

\smallskip

Note here that, in general,
a physical background ``talks'' with the fluctuations, and it is this coupling that gives rise to an instability of the constant--curvature backgrounds as we discuss below.
The essential effect of backreaction models is not a large magnitude of $\CQ_{\CD}$, but a 
dynamical coupling of a nonvanishing $\CQ_{\CD}$ to the averaged scalar curvature deviation $\CW_\CD$. This implies that the temporal
behavior of the averaged curvature deviates from the behavior of a constant--curvature model. In concrete studies, as discussed further below, this turns out to be the major effect of backreaction, since it does not only change the kinematical properties of the cosmological model, but also the interpretation of observational data as we explain in Section~IV.

\subsection{Scalar field emerging from geometrical inhomogeneities}

We rewrite the above set of spatially averaged equations together with their integrability conditions by appealing to the kinematical equations of the standard model, which will now be sourced by an effective perfect fluid energy--momentum tensor \cite{buchert:fluid}:
\begin{eqnarray}
&3\frac{{\ddot{a}}_{\CD}}{a_{\CD}}  =  -4\pi G(\varrho_{\rm eff}^{\CD}+3{p}_{\rm eff}^{\CD})\;+\;\Lambda \;\;\;\;;\\
&3H_{\CD}^{2}- \frac{3 k_\CD}{a_\CD^2}= 8\pi G\varrho_{\rm eff}^{\CD}\;+\;\Lambda
\;\;\;\;;\\
&{\dot{\varrho}}_{\rm eff}^{\CD}+3H_{\CD}\left(\varrho_{\rm eff}^{\CD}+{p}_{\rm eff}^{\CD}\right)\;=0\;,
\label{eq:effectivefriedmann}
\end{eqnarray}
where the effective densities are defined as 
\begin{eqnarray}
\varrho_{{\rm eff}}^{{\CD}} := \average{\varrho} + \varrho_{\Phi}\;\;\;;\;\;\varrho_{\Phi} & := & -\frac{1}{16\pi G}{\CQ}_{{\CD}}-\frac{1}{16\pi G}\CW_\CD \;;\nonumber
\label{eq:equationofstate}\\
{p}_{{\rm eff}}^{{\CD}} := p_{\Phi}\;\;\;;\;\;\;p_{\Phi}& := & -\frac{1}{16\pi G}{\CQ}_{{\CD}}+\frac{1}{48\pi G}\CW_\CD \;.
\end{eqnarray}
In this form the effective equations suggest themselves to interpret the extra fluctuating sources in terms of a scalar field \cite{buchert:static,morphon}, which refer to the inhomogeneities in geometrical variables.
Thus, we choose to consider the averaged model as a (scale--dependent) ``standard model'' with matter source evolving in a {\it mean field} of backreaction terms.
This scalar field we call the {\it morphon field}, since it captures the morphological (integral--geometrical \cite{buchert:review}) signature of structure. (Note that in more general cases, involving lapse and shift functions, the structure of the scalar field theory suggested by the equations may no longer be a minimally coupled one.) We rewrite \cite{morphon}:
\begin{equation}
\label{morphon:field}
\varrho^\CD_{\Phi}=\epsilon \frac{1}{2}{{\dot\Phi}_\CD}^2 + U_\CD\;\;\;;\;\;\;p^\CD_{\Phi} =
\epsilon \frac{1}{2}{{\dot\Phi}_\CD}^2 - U_\CD\;\;,
\end{equation}
where $\epsilon=+1$ for a standard scalar field (with positive kinetic energy), and 
$\epsilon=-1$ for a phantom scalar field (with negative kinetic energy; if $\epsilon$ is negative, a ``ghost'' can formally arise on the level of an
effective scalar field, although the underlying theory does not contain one; note also that there is no violation of energy conditions, since we have only dust matter).
Thus, from the above equations, we obtain the following correspondence:
\begin{equation}
\label{correspondence1}
-\frac{1}{8\pi G}{\CQ}_\CD \;=\; \epsilon {\dot\Phi}^2_\CD - U_\CD\;\;\;;\;\;\;
-\frac{1}{8\pi G}\CW_\CD = 3 U_\CD\;\;.
\end{equation} 
The correspondence (\ref{correspondence1}) recasts the integrability conditions (\ref{eq:integrability})
into a (scale--dependent) Klein--Gordon equation for $\Phi_\CD$, and 
${\dot\Phi}_\CD \ne 0$:
\begin{equation}
\label{kleingordon}
{\ddot\Phi}_\CD + 3 H_{\cal D}{\dot\Phi}_\CD + 
\epsilon\frac{\partial}{\partial \Phi_\CD}U(\Phi_\CD , \average{\varrho})\;=\;0\;\;.
\end{equation} 
We appreciate that the deviation of the averaged scalar curvature from a constant--curvature model is directly proportional to the potential energy density of the scalar field.
Averaged universe models obeying this set of equations follow, thus, a Friedmannian kinematics with a fundamental matter source, and an effective scalar field source that reflects the shape of spatial hypersurfaces and the shape of their embedding into spacetime. 
Given the potential in terms of variables of the averaged system, the evolution of these models is fixed (the governing equations are closed). This also potentially fixes coupling parameters,
since all involved fields can be traced back to the initial value problem of general relativity.

\smallskip

The morphon formulation of the backreaction problem opens a nice interpretation in terms of energies: a homogeneous model, $\CQ_\CD = 0$ (a necessary and sufficient condition to also drop the scale--dependence, if required on every scale), is characterized by the {\it virial equilibrium condition}:  
\begin{equation}
\label{virialbalance}
2\, E^\CD_{\rm kin} \,+\, E^\CD_{\rm pot} =\;- \frac{\CQ_\CD V_\CD}{8\pi G}\;\;\;\;,\;\;\;\;\CQ_\CD = 0 \;\;\;\;;\;\;\;\;E^\CD_{\rm kin} = \varepsilon{\dot\Phi}_\CD^2 V_\CD \;\;\;,\;\;\;\;
E^\CD_{\rm pot} = -U_\CD V_\CD \;\;.
\end{equation} 
Deviations from homogeneity, $\CQ_\CD \ne 0$, thus invoke a non--equilibrium dynamics of the morphon in its potential that is dictated by the effective intrinsic curvature of the space in which the fluctuations evolve. Morphon energies are redistributed and can be assigned to the {\it dark energies}. Dependent on the signs of the backreaction terms (and a sign change may occur by looking at different spatial scales) the morphon can act as a scalar field model for {\it Dark Matter}, a quintessence model for {\it Dark Energy}, or it can even play the role of a {\it classical inflaton}, as we exemplify in the following subsection. (For the different interpretations of scalar fields see the review \cite{DE:review}, and for unified views the selection of papers \cite{arbey,matos:unification,sahni,chaplygin}, and for scalar Dark Matter e.g. \cite{matos:dm, matos:halos,matos:dm09}.

\subsection{Example: morphonic inflation}

Consider a tube of space time characterized by a gravitational field with no fundamental sources. The $4-$Ricci curvature tensor vanishes everywhere, but not necessarily the $4-$Weyl curvature tensor.
Even if this classical vacuum space time is initially foliated into $3-$Ricci flat hypersurfaces, this does not remain so in the dynamical evolution: such an initially prepared homogeneous state is unstable (a fact that we shall explain in the next section), and these hypersurfaces, if perturbed, will necessarily develop into inhomogeneous hypersurfaces featuring non--vanishing averaged curvature invariants, i.e. an intrinsic, on average negative curvature, and a compensating extrinsic curvature due to the embedding of the hypersurfaces into the Ricci--flat space time. Thus, in this picture, the space section will develop a morphonic scalar field that is driven by a Klein--Gordon dynamics and specified by the initial value of its self--interaction potential. 
While this instability is dynamical, the picture reminds us of the behavior of a fundamental inflaton, where the instability is created by an externally added potential. 

\smallskip

We specify initial data according to the analogy of the backreaction variables to the morphon field ($\CQ_\CD^i  \equiv \CQ_\CD (t_i);  \CW_\CD^i  \equiv \CW_\CD (t_i)$):
\begin{equation}
U^i_\CD \equiv -\frac{1}{24\pi G}\CW_\CD^i \;\;\;;\;\;\; {\dot\Phi}_\CD^i \equiv  \sqrt{\frac{-\CQ_\CD^i}{8\pi G\epsilon} + U^i_\CD } \;\;\;;\;\;\;\Phi_\CD^i \equiv \Phi_\CD (t_i )\;.
\label{const-pot}
\end{equation}
Interestingly, for a homogeneous initial state, $\CQ_\CD (t_i) = 0$, the kinetic energy density of the morphon field is initially non--vanishing, and the Klein--Gordon dynamics drives the morphon into a stable fix point, an (assumed) existing minimum of the potential. However, the outcome does not depend much on the initial data for $\CQ_\CD$: we could also start with inhomogeneous initial data, e.g. a cosmological constant that is mimicked by a particular morphon, in which case the initial kinetic energy density is zero. Curvature energy is thus converted into kinetic energy, driving the system into an accelerated expansion phase. The value of the potential is necessarily always positive, since the vacuum $3-$space has negative intrinsic curvature. The inflationary mechanism is thus the same as the mechanism to create an accelerated expansion in a Quintessence phase.

\begin{figure}
\includegraphics[width=0.6\textwidth]{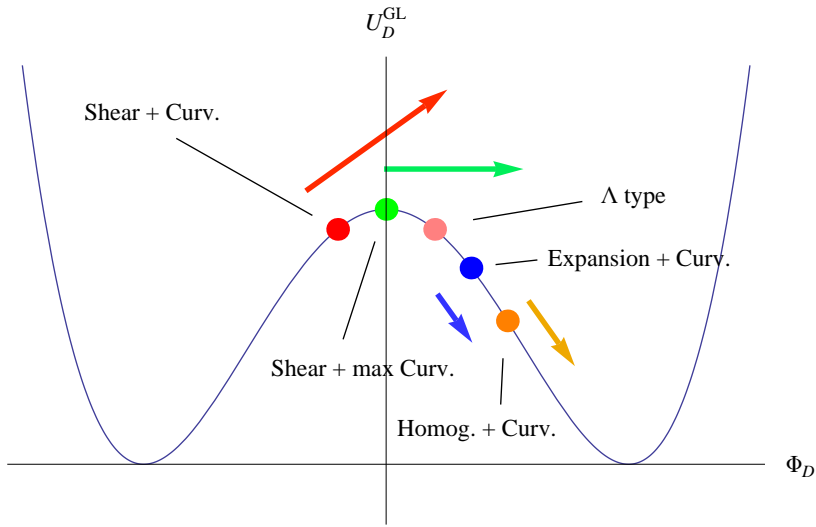}
\caption{The Ginzburg--Landau potential in arbitrary units and the possible 
initial conditions as well as their physical meaning.
All conditions possess some curvature $\CW_\CD^i  < 0$.
The arrows schematically indicate the amplitude of the morphon's initial speed $\dot\Phi_\CD^i$. 
In the order of the points (from left to right): the first two points dominated by shear fluctuations (red, green) are obtained for 
$\CQ_\CD^i < 0 \Leftrightarrow \dot\Phi_\CD^{i\:2} > 2(H_\CD^{i\:2}+ k_\CD^i )$; 
the next points dominated by expansion fluctuations (blue, pink) for $ \dot\Phi_\CD^{i\:2} < 2(H_\CD^{i\:2}+ k_\CD^i )$,
where the de Sitter--$\Lambda$ equivalent case has a stiff morphon $\dot\Phi_\CD^i = 0$; 
the homogeneous case (last point, orange) is obtained for  $ \dot\Phi_\CD^{i\:2} = 2(H_\CD^{i\:2}+ k_\CD^i )$. Figure from \cite{buchertobadia:inflation}.\label{fig:GLpotential}}
\end{figure}

We realize the inflationary scenario \cite{buchertobadia:inflation} by closing the set of averaged equations with a potential of the generic Ginzburg--Landau form: 
\begin{equation}
\label{GLpotential}
U_\CD^{GL} = U_0 \left(\Phi_\CD^2 - \Phi_0^2 \right)^2/\Phi_0^4 \;.
\end{equation}
The position of the minimum $\Phi_0$ and the amplitude $U_0$ 
play different roles: the first one fixes the duration of inflation,
while the second sets the size of the Hubble radius at which it happens. This potential has been extensively studied in the context of chaotic inflation \cite{Linde:1983gd}. The various initial conditions together with their interpretation in terms of geometrical properties of space are shown in Figure~\ref{fig:GLpotential}. 

Combining this purely morphonic picture of inflation created from the Einstein vacuum with a fundamental scalar field, we can establish hybrid inflationary models with two scalar fields, one of them being the morphon that is always present in the case of inhomogeneous universe models. 

\section{Global gravitational instability of the standard model background}

\subsection{The phase space of exact background states}

\begin{figure}
\includegraphics[width=0.525\textwidth]{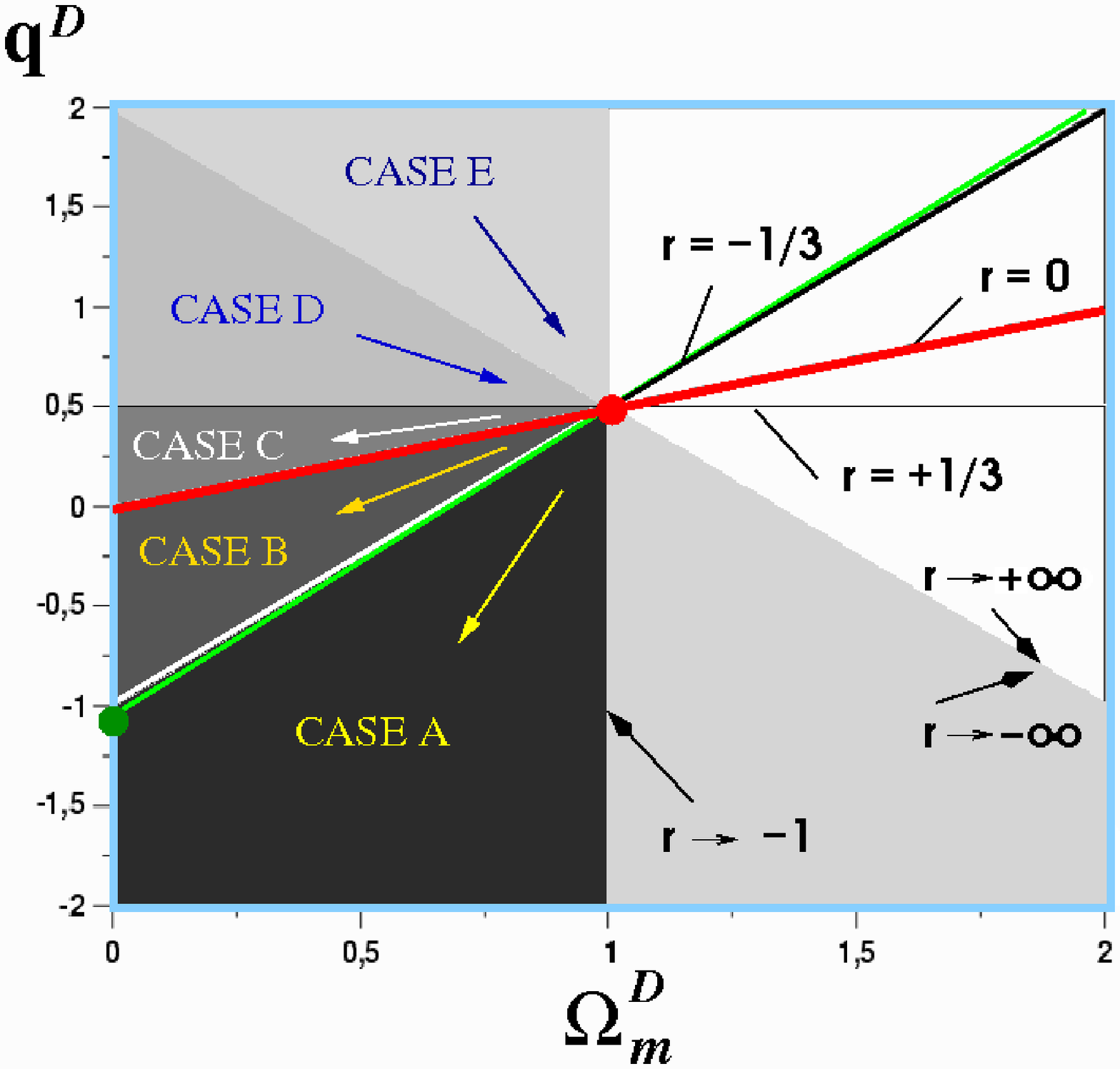}%
\hspace{0.05\textwidth}%
\includegraphics[width=0.415\textwidth]{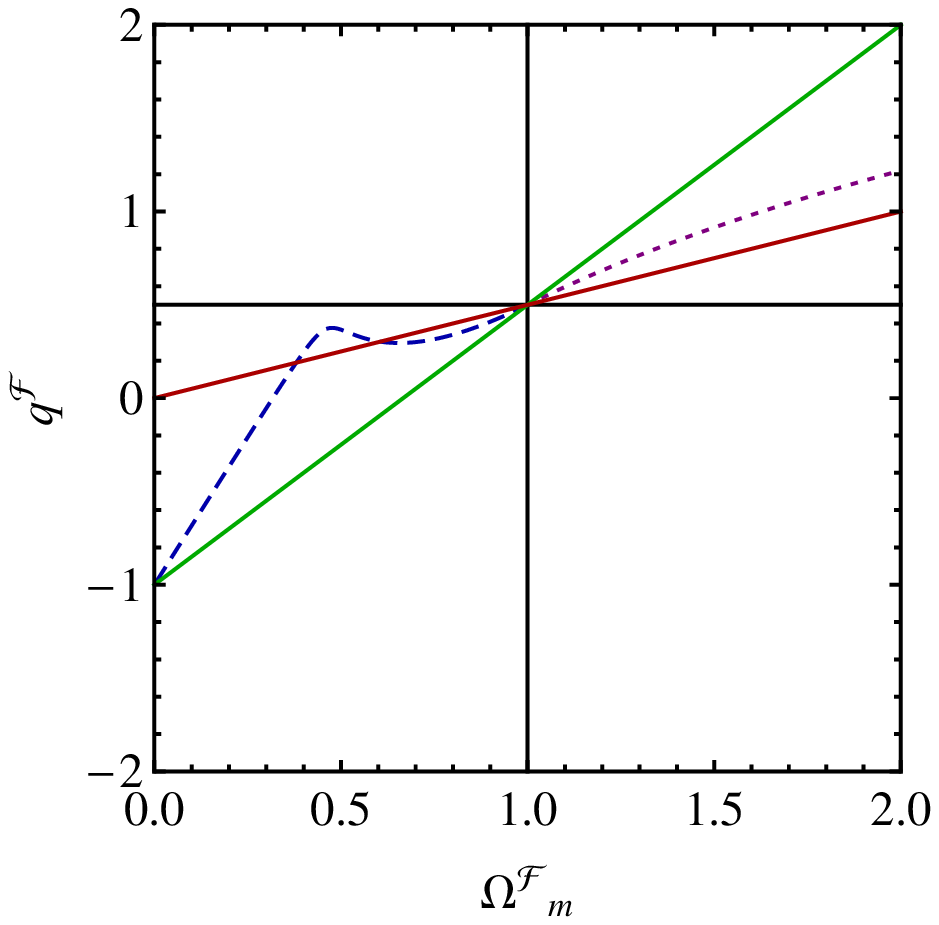}
\caption{{}Left: ``Cosmic phase space'' of the solutions of the averaged equations (``generalized backgrounds'') in a plane spanned
by the volume deceleration parameter $q^\CD := -{\ddot a}_\CD / (a_\CD H_\CD^2) = 1/2
\Omega_m^{\CD} + 2 \Omega_{\CQ}^{\CD} - \Omega_{\Lambda}^{\CD}$ and the matter density parameter \cite{morphon}.
It represents a two--dimensional subspace $\lbrace\;\Lambda =0\;\rbrace$  of the full solution space that would include a cosmological constant.
The segments are separated by particular exact scaling solutions of the full problem. We identify the following scaling solutions:
all the scaling solutions are represented by straight lines passing
through the Einstein--de Sitter model in the center of the diagram ($1/2;1$). 
Models with ``Friedmannian'' kinematics, but with renormalized parameters form the line 
$r=1/3$ (for details see \cite{morphon}, Appendix A). The line $r=0$ are models with no backreaction 
on which the parameter $\Omega^\CD_k$ varies
(scale--dependent ``Friedmannian models'').
Below the line $r=0$ in the ``quintessence phase''
we find effective models with subdominant shear fluctuations
(${\CQ}_\CD$ positive, $\Omega^\CD_{\CQ}$ negative). The line $r=-1/3$
mimics a ``Friedmannian model'' with cosmological constant.
The line below $r=-1/3$ in the ``phantom quintessence phase''
represents the solution inferred from SNLS data ({\it cf.} \cite{morphon}),
and the point at ($q^{\CD};\Omega^\CD_m) = (-1.03; 0)$ 
locates the late--time attractor associated with this solution. Since we have no cosmological constant here, all expanding solutions in the subplane $q^\CD <0$  drive the averaged variables away from the standard model featuring a backreaction--driven volume acceleration of effectively isotropic cosmologies that are curvature--dominated at late times.\\
Right: we show the evolution of phase space orbits (running away from the standard model) for a multiscale model that is explained further below; this model is partitioned into over--dense $\CM$ (dotted) and under--dense $\CE$ (dashed) regions, their volume fraction being derived from $N$--body simulations \cite{multiscale}. They
are shown here in the same plot for economic reasons and actually live in two different phase spaces corresponding to large scales for the $\CE$--regions and small scales for the $\CM$--regions. $\CF$ denotes one of the regions $\CM$ or $\CE$.}
\label{fig:phasespace}
\end{figure}

The space of possible states of an averaged cosmological model, or the space of ``generalized backgrounds'' has one dimension more than the space of possible homogeneous--isotropic solutions in the standard model framework. This can be seen by introducing adimensional ``cosmological parameters''.   
We divide the volume--averaged expansion law 
by the squared {\em volume Hubble functional} $H_\CD := {\dot a}_\CD / a_\CD$ introduced before.
Then, the expansion law can be expressed as a sum of adimensional average characteristics:
\begin{equation}
\label{omega}
\Omega_m^{\CD}\;+\;\Omega_{\Lambda}^{\CD}\;+\;\Omega_k^\CD \;+\;\Omega_{\CW}^{\CD}\;+\;
\Omega_{\CQ}^{\CD}\;=\;1\;\;,
\end{equation}
with:
\begin{eqnarray}
&\Omega_m^{\CD} : = \frac{8\pi G}{3 H_{\CD}^2} \langle\varrho\rangle_{\cal D}  \;\;;\;\;
\Omega_{\Lambda}^{\CD} := \frac{\Lambda}{3 H_{\CD}^2 }\;\;;\;\;
\Omega_{k}^{\CD} := - \frac{k_{\initial\CD}}{a_\CD^2 H_{\CD}^2 }\;\;\;;\nonumber\\
&\Omega_{\CW}^{\CD} := - \frac{\CW_\CD}{6 H_{\CD}^2 }\;\;;\;\;
\Omega_{\CQ}^{\CD} := - \frac{{\CQ}_{\CD}}{6 H_{\CD}^2 } \;\;.
\end{eqnarray}
Taking for simplicity the constant--curvature parameter and the curvature deviation into a single full curvature parameter, $\Omega_k^\CD + \Omega_\CW^\CD = :\Omega_\CR^\CD$, the generalized model offers a
{\it cosmic quartet} of parameters. Furthermore, if we put $\Lambda =0$, the expansion law defines, for each scale, a two--dimensional phase space of states. A one--dimensional subset of this phase space is formed by ``backgrounds'' with Friedmannian kinematics (see Figure~\ref{fig:phasespace}). 

\smallskip

We can analyze the fix points and their stability properties in the general dynamical system \cite{morphon}, \cite{phasespace}.
The principal outcome of this study is that the standard zero--curvature model forms a {\it saddle point}; of particular interest are two instability sectors for the standard model, regarded as averaged state:
firstly, perturbed homogeneous states are driven into a sector of highly isotropic, negative curvature and accelerated expanding ``backgrounds'' where backreaction thus mimics Dark Energy behavior over the domain $\CD$; secondly,  perturbed homogeneous states are driven into a sector of highly anisotropic, positive curvature, collapsing and decelerated ``backgrounds'' where backreaction thus mimics Dark Matter behavior over the domain $\CD$. Concrete models show that the former happens on large scales, and the latter on the scales of galaxy surveys, and also on smaller scales.
Thus, qualitatively, the instability sectors identified comply with the aim to trace the dark components back to physical properties, but they also agree with the expected properties of the structure: isotropic, accelerating states on large scales, and highly anisotropic structures on the filamentary distribution of superclusters. Moreover, the curvature properties also meet the expectations: on large scales the Universe is void--dominated and, hence, dominated by negative curvature, while on intermediate scales over--densities are abundant and are characterized by positive curvature.

\smallskip


\subsection{Dark Energy and Dark Matter hidden in the geometry of space}

The fact that the standard model is globally unstable in the phase space of averaged states, and the fact that the instability sectors lie in the right corners to explain Dark Energy and Dark Matter behavior, are both strong qualitative arguments to expect that the conservative explanation of the dark energies through morphon energies is valuable. The underlying mechanism is indeed based on the fundamental existence of the relation between geometrical curvatures and sources dictated by Einstein's equations. 

\smallskip

Whether this mechanism is sufficient in a quantitative sense is to date still an open issue. The difficulty to construct quantitative models is to be seen in the need for non--standard tools, for example perturbation theory on a fixed reference background should be replaced by a fluctuation theory on evolving backgrounds that captures the average over the fluctuations.
The question whether perturbations are small can only be answered if we know with respect to which background they are small. Furthermore, since backreaction affects the geometry, it will change the interpretation of observational data, a problem that is intimately related to the generalization of the cosmological model, and to which we shall come below. 

\smallskip

Before, we shall in the next section explain the identified mechanism by discussing some physical properties of structure formation and its relation to the interpretation of geometrical curvature invariants and how they mimic the dark sources. We here touch on a deeper problem: backreaction effects account for both, Dark Energy and Dark Matter, simultaneously. Whether, on a given domain, or on an ensemble of domains on a given scale, the  morphon mimics Dark Energy or Dark Matter behavior,  changes as a function of time and as a function of scale. Moreover, the small--scale contribution to e.g. a Dark Matter behavior requires more sophisticated relativistic models than the dust model used throughout here (e.g. \cite{dm1}, \cite{dm2}). Considering rotation curves of galaxy halos, missing gravitational sources in clusters or missing sources on cosmological scales  always needs different modeling strategies. We try in the following to provide a first step of disentangeling Dark Energy and Dark Matter behavior by explicitly constructing an effective multiscale cosmological model.  

\subsection{Multiscale cosmology and structure--emerging volume acceleration}

Contrary to the standard model, where a homogeneous background is used as a standard of reference for the expansion history of the Universe, a background constructed as the average over fluctuating fields introduces a subtle element: while a homogeneous geometry can be characterized locally, an average is nonlocal, since it is determined by the inhomogeneities inside, but also outside the averaging domain, reflecting the nonlocal nature of gravitation. Furthermore, an average incorporates correlations of the local fields. It is this latter which is the key--driver of a repulsiv ``effective pressure'' that arises in the averaged models. 

\smallskip

This ``effective pressure'' provides the reason why backreaction can produce a volume--accelerating component despite the decelerating nature of the general local acceleration law. This can be seen easily by comparing the local and the volume--averaged Raychaudhuri equation (for vanishing vorticity and pressure that both would also act accelerating on the local level, but only on small scales):
\begin{equation}
\label{raychaudhuri}
\dot{\theta} = \Lambda - 4\pi G \varrho + 2{\rm II} - {\rm I}^2 \;\;\;\;;\;\;\;\; \langle\theta\dot\rangle = \Lambda - 4\pi G \average{\varrho} + 2\average{{\rm II}} - 
\average{\rm I}^2\;\;,
\end{equation}
where we defined the principal scalar invariants of the expansion tensor $\Theta_{ij}$, $2{\rm II}:= 2/3 \theta^2 - 2\sigma^2$ and ${\rm I}:=\theta$.

\smallskip

Clearly, by shrinking the domain to a point, both equations agree. However, evaluating the local and averaged invariants,
\begin{eqnarray}
& 2{\rm II}- {\rm I}^2 = -\frac{1}{3} \theta^2 -2\sigma^2\;\;\;\;;\nonumber\\
& 2\average{\rm II} - \average{\rm I}^2 = 
\frac{2}{3} \average{(\theta - \average{\theta})^2} - 2 \average{(\sigma - \average{\sigma})^2} - \frac{1}{3} \average{\theta}^2 - 2\average{\sigma}^2\;\;,
\end{eqnarray}
gives rise to two additional, positive--definite fluctuation terms, where that for the averaged expansion variance enters with a positive sign.
Thus, the time--derivative of a (on some spatial domain $\CD$) averaged expansion may be positive despite the fact that the time--derivative of the expansion {\em at all points} in $\CD$ is negative. 

\smallskip

In concrete models this variance is the source of a possible large--scale volume--acceleration that would be assigned to Dark Energy in the standard model, while the averaged shear fluctuations mimic an attractive source that would be missing as Dark Matter in the standard model on cosmological scales. Both terms are competing in the kinematical backreaction $\CQ_\CD$.
Since backreaction depends on scale, it may act in both ways.

\smallskip

We can go one step further and make the scale--dependence explicit by introducing a union of disjoint over--dense regions $\CM$ and a union of disjoint under--dense regions
$\CE$, which both make up the total (homogeneity--scale) region $\CD$. The averaged equations can be split accordingly yielding for the kinematical backreaction \cite{buchertcarfora}, \cite{multiscale}:
\begin{equation}
\CQ_{\CD} \; =\;  \lambda_{\CM}\CQ_{\CM}+\left(1-\lambda_{\CM}\right)\CQ_{\CE}
+6\lambda_{\CM}\left(1-\lambda_{\CM}\right)\left(H_{\CM}-H_{\CE}\right)^{2}\;,
\end{equation}
where $\lambda_{\CM}:=\left|\CM\right|/\left|\CD\right|$ denotes the volume--fraction of the over--dense regions compared to the volume of the region $\CD$. In a Gaussian random field this fraction would be $0,5$ and would gradually drop in a typical structure formation scenario that clumps matter into small volumes and that features voids that gradually dominate the volume in the course of structure formation.

\smallskip

\begin{figure}
\includegraphics[width=0.75\textwidth]{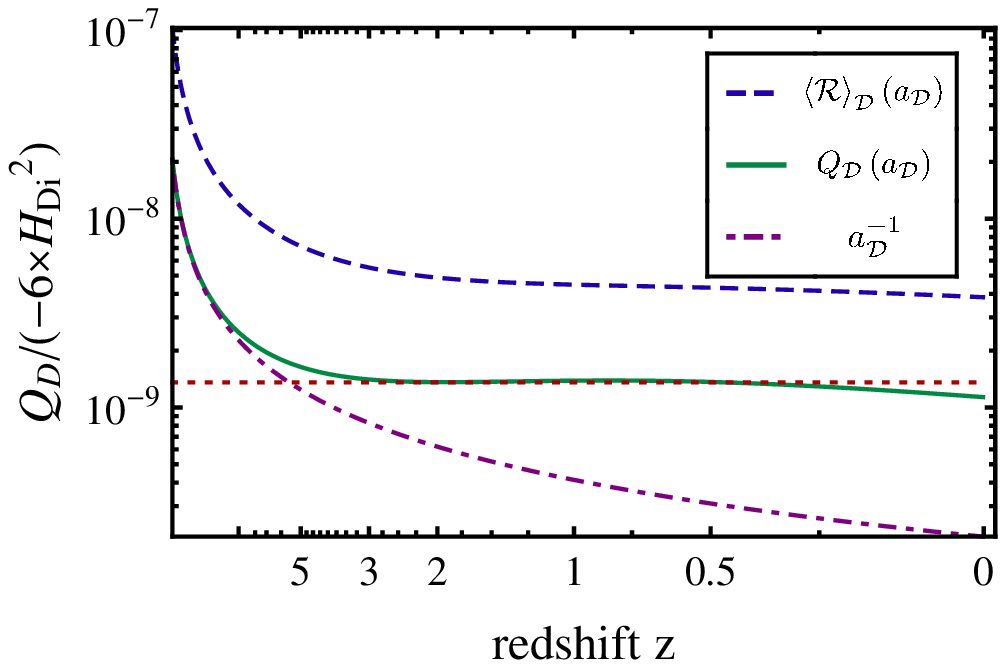}
\caption{Plot of the evolution of $\CQ_{\CD}$ and $\average{\CR}$ in terms of the global scale factor $a_{\CD}$. For comparison a line with a simple $a_{\CD}^{-1}$--scaling is added, that corresponds to the leading mode in second--order perturbation theory, and one that is constant. $\CQ_{\CD}$ and $\average{\CR}$ are normalized by $-6H_{\CD_{\rm{i}}}^{2}$ so that the values at the initial time represent $\Omega_{\CQ}^{\CD_{\rm{i}}}$ and $\Omega_{\CR}^{\CD_{\rm{i}}}$. We appreciate that the backreaction terms feature an approximate cosmological constant behavior on the homogeneity scale despite the assumption of $a_{\CF}^{-1}$--scaling on the partitioned domains $\CF = \CM , \CE$. Physically, this result can be attributed to the expansion variance between the subdomains and, hence, this latter is identified as the key effect to produce a global Dark Energy--like behavior of the backreaction terms. Figure from \cite{multiscale}.
\label{fig:multiscale}}
\end{figure}%

Ignoring for simplicity the individual backreaction terms on the partitioned domains, the total backreaction features a positive--definite term that describes the variance between
the different expansion histories of over-- and under--dense regions. It is this term that generates a Dark Energy behavior over the domain $\CD$ (see also \cite{rasanen:peakmodel} for a model by R\"as\"anen, and \cite{wiltshire:clocks,wiltshire:solutions,wiltshire:timescape,wiltshire:obs} for Wiltshire's model that is based on this term only, but includes a phenomenlogical lapse function to account for different histories in $\CM$ and $\CE$ regions that, this latter, we cannot implement in the synchronous foliation of a multiscale dust model). If we model non--zero individual backreaction terms by an extrapolation of the leading perturbative mode in second--order perturbation theory \cite{gaugeinv,li:scale} that also corresponds to the leading order in a Newtonian non--perturbative model \cite{bks}, then we even produce a cosmological constant behavior over $\CD$, see Figure~\ref{fig:multiscale}.
In other words, the fact that, physically, over--dense regions tend to be gravitationally bound, i.e. do not partake significantly in the global expansion, already produces a large--scale ``kinematical pressure'' as a source of volume acceleration. A homogeneous background simply cannot account for this difference.

\newpage

\section{Inhomogeneous and effective metrics}

\subsection{Some notes on relativistic perturbation theories}

Consider a spatial metric form $\bf g$ with coefficients $g_{ij}$ in an exact (co--tangential) basis ${\bf d}X^i \otimes {\bf d}X^j$. We can write any metric as a quadratic form of deformation one--forms, ${\bf g} = \delta_{ab} \boldsymbol{\eta}^a \otimes \boldsymbol{\eta}^b$, i.e. in terms of coefficients, $g_{ij} = \delta_{ab} \eta^a_{\;\,i}\eta^b_{\;\,j}$. Now, such a metric form is {\it homogeneous}, i.e. its Ricci tensor vanishes everywhere, if there exist functions $f^a$, such that the one--forms can be written as exact forms, $\boldsymbol{\eta}^a \equiv {\bf d} f^a$. In other words, if we can find a coordinate transformation $x^i = f^{a\equiv i} (X^j , t)$ that transforms the Euclidean metric coefficients in a new basis,  ${\bf d}x^i \otimes {\bf d}x^j$, $\delta_{ij} dx^i dx^j = \delta_{ab} f^a_{\;\,|i} f^b_{\;\,|j} dX^i dX^j$, with a vertical slash denoting partial spatial derivative, into the metric coefficients $g_{ij}$, then these latter are just a rewriting of the homogeneous space. Given this remark, any perturbation theory that features metric forms of the integrable form, does not describe relativistic inhomogeneities; metric coefficients of the form $g_{ij} = \delta_{ab} f^a_{\;\,|i} f^b_{\;\,|j}$ describe Newtonian (Lagrangian) perturbations on a flat background space. A truely relativistic perturbation theory deforms the background geometry, in other words, the perturbations live in a perturbed space, not on a reference background. This remark also shows that relativistic perturbation terms can never contain full divergences, since this latter needs integrable one--form fields. 

\smallskip

In light of these introductory remarks, an inhomogeneous relativistic metric produces curvature that, if volume--averaged on some domain, does not obey a conservation law (as can be explicitly seen in the coupling equation to the fluctuations (\ref{eq:integrability})) in the sense that it would always average out to zero; for details on curvature estimates see \cite{buchertcarfora}). This fact in itself shows the existence of a dynamical evolution of an averaged curvature, as soon as structures form. On the contrary, standard perturbation theory formulated on a fixed flat background is constructed such that the averages always vanish on the background, demonstrating the limited nature of results obtained by standard perturbative arguments.

\smallskip

Another perturbative argument aims to justify the validity of the homogeneous geometry, even down to the scales of neutron stars \cite{ishibashi}, since perturbations of the metric remain small with respect to the flat background. This argument does not contradict the existence of a large backreaction effect, since these latter depend on first and second derivatives of the metric \cite{estim}, \cite{kolb:voids}, \cite{rasanen:perturbation}. Also, the perturbations are considered on a flat background that does not interact with structure. As we explained in detail, the perturbations may be small on a different (physical) background, in which case a perturbation may already live in a background with strong 
curvature (a zero--order effect). It is therefore not fruitful to argue against the relevance of backreaction within standard limited schemes, but rather an effort to generalize perturbation theory is needed.

\subsection{Template metrics and effective distances}

For the construction of an effective cosmological evolution model, as outlined above, a metric needs not be specified.  The need for the construction of an effective metric in these models arises, since measured redshifts have to be interpreted in terms of distances along the light cone. Given an explicit, generic and realistic, inhomogeneous metric, the need for the construction of effective metrics does not arise.
Also, if we succeed to understand the evolution of light cone averages in relation to distances, then also here an explicit metric will not be needed
\cite{rasanen:light,rasanen:light2}, \cite{lightcone}.

\smallskip

The idea of an effective cosmological metric comes from the ``fitting problem'', that has been particularly emphasized by George Ellis already in the 70's \cite{ellis:average}. The observation was that an inhomogeneous metric does not average out to a homogeneous 
metric that forms a solution of general relativity. Not only the nonlinearity of the theory, but also simple arguments of a non--commutativity \cite{ellisbuchert} between evolution equations and the averaging operation, give rise to the need to find a ``best--fit'', we may call it ``template'' geometry, that inherits homogeneity and (almost--)isotropy on the large scales and, at the same time, incorporates the inhomogeneous structure ``on average'' (see also the early practical implementations of this problem \cite{futamase1,futamase2}, \cite{ellisstoeger}, \cite{hellaby:volume}).

\smallskip

For the solution of the {\it fitting problem} various strategies have been proposed (see \cite{ellisbuchert} and references therein). One strategy, that allows to explicitly perform a ``smoothing'' of an inhomogeneous metric into a constant--curvature metric at one instant of time, is based on Ricci--flow theory: one notices that a smoothing operation of metrical properties can be put into practice by a {\it rescaling} of the metric in the direction of its Ricci curvature. The scaling equations for realizing this are well--studied, and the rescaling flow results in a constant--curvature metric that carries ``dressed'' cosmological variables \cite{klingon}, \cite{dressing}. These incorporate intrinsic curvature backreaction terms describing the difference to the ``bare'' cosmological parameters as they are obtained through kinematical averaging. 

\subsection{Reinterpretation of observational data}

The standard method of interpreting observations is to construct the light cone $ds^2 = 0$ from the line--element $ds^2 = -dt^2 + g^{\rm hom}_{ij} dX^i dX^j$, where the coefficients $g^{\rm hom}_{ij}$ are given in the form of a constant--curvature (FLRW) metric, and then to calculate the luminosity distance $d_L (z)$ in this metric for a given observed redshift $z$. Assuming this metric for the inhomogeneous Universe implies the conjecture that the FLRW metric is the correct ``template'' of an effective cosmological metric. However, the integrated exact equations (the integral properties of a general inhomogeneous model) are not compatible with this metric, simply because the averaged curvature is assumed to be of the form
$\average{\CR} = 6 k a^{-2}$ on all scales. Improving the metric template slightly, by replacing the global scale factor $a(t)$ through the volume--scale factor $a_\CD (t)$ and the integration constant $k$ through the domain--dependent integration constant $k_\CD$, renders this metric implicitly scale--dependent \cite{singh2}. As we explained, this is not enough since the averaged curvature couples to the inhomogeneities and in general deviates from the $a_\CD^{-2}$--behavior. What we can do as a first approximation, and this would render the metric compatible with the kinematical average properties, is to introduce the exact averaged curvature in place of the constant curvature in this metric form \cite{morphon:obs}.

\smallskip

The resulting effective space time metric consists of a synchronous foliation of constant--curvature metrics that are, however, parametrized by the exact integral properties of the
inhomogeneous curvature, thus they ``repair'' the standard template metric as for the evolution properties of spatial variables. Such a construction can be motivated by Ricci--flow smoothing, that guarantees the existence of smoothed--out constant curvature sections at one instant of time, and by assuming that the intrinsic backreaction terms are subdominant, so that we can parametrize the metric by ``bare'' kinematical averages. To stack these hypersurfaces together introduces, however, an inhomogeneous light cone structure \cite{mersinik(t)}, \cite{rasanenk(t)}. Ideally, one would wish to smooth the light cone too, which is also possible by employing Ricci flow techniques. Improving this first approach to a template metric is needed and this is work in progress.

\smallskip

\begin{figure}
\includegraphics[width=0.45\textwidth]{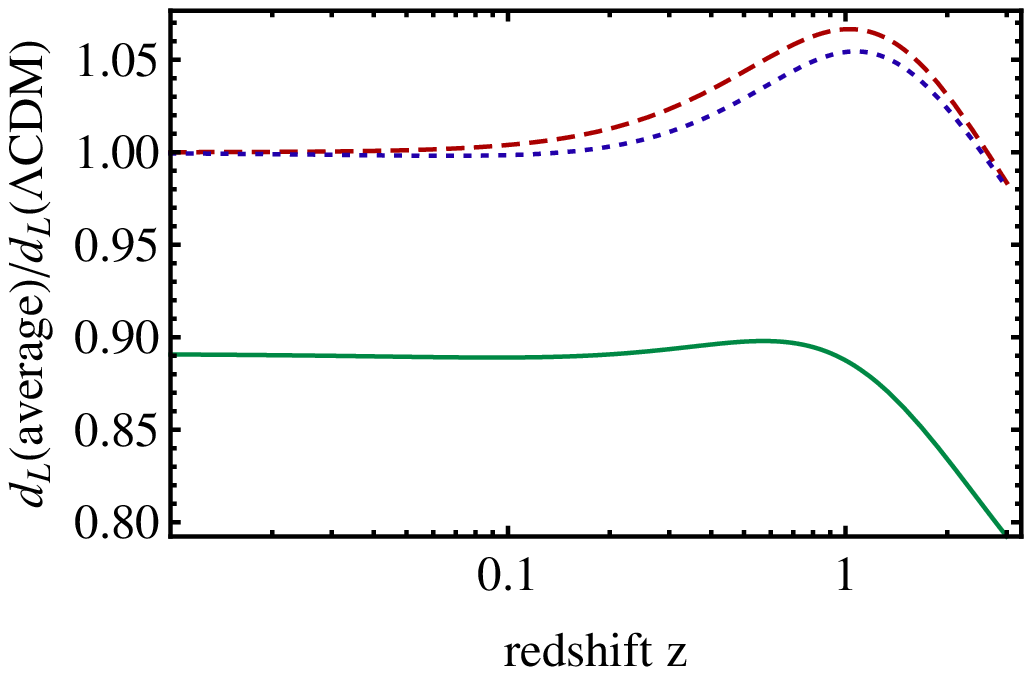}%
\hspace{0.04\textwidth}%
\includegraphics[width=0.45\textwidth]{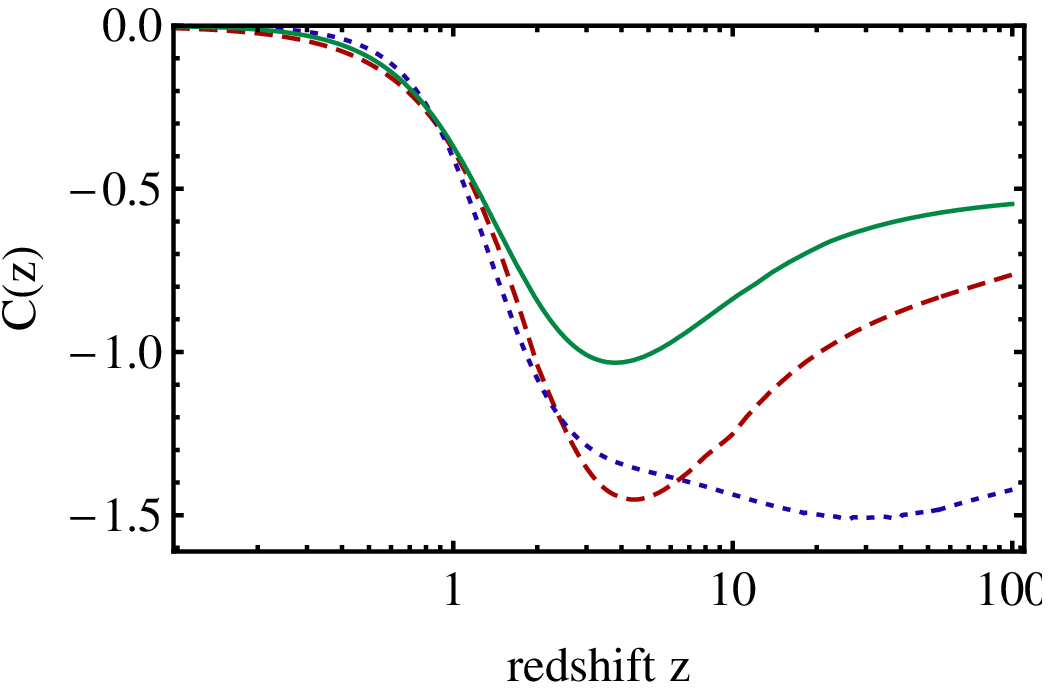}
\caption{Left: Comparison of the luminosity distances of the multiscale models investigated in \cite{multiscale} and based on the template metric of \cite{morphon:obs}, with the one of a flat $\Lambda$CDM model with $h=0.7$ and $\Omega_{m}=0.27$. On top a model where we force the volume scale factor $a_{\CD}$ to follow the $\Lambda$CDM evolution. Despite this assumption, the changing curvature affects the luminosity distance. The luminosity distances in these models show a significant feature at a redshift of around $1$, when compared with the best fit $\Lambda$CDM model, which may be looked for in the SN data. The curve below is a model with $a_\CD^{-1}$--scaling. For comparison we also included the luminosity distance of the best fit model of \cite{morphon:obs}. Because of a different Hubble rate of $h=0.7854$ it lies below the others from the beginning. This model does not significantly show the distinct feature of the other two models around a redshift of $1$, due to the assumption of a single--scale cosmology.\\Right: Values of Clarkson's $C$--function \cite{clarkson} for the best fit model of \cite{morphon:obs} (top), the model where the scale--factor is forced to follow the $\Lambda$CDM evolution (middle), and the model with $a_\CD^{-1}$--scaling (bottom). Recall that, for every Friedmann model, $C(z)$ vanishes exactly on all scales and for all redshifts. For the inhomogeneous models shown in the plot, this function has a minimum which may serve as observational evidence for the effective cosmologies, as proposed in \cite{morphon:obs}. As both multiscale models show, it is not even necessary to measure derivatives of distance, since the feature is already present in the distance. \\
Figure from \cite{multiscale}.\label{fig:distance}}
\end{figure}

The result of employing the improved template metric described above is a change in the luminosity distance that would alter the interpretation of all observational data formerly based on FLRW distances. Examples for the multiscale models investigated in \cite{multiscale} are presented in Figure~\ref{fig:distance}.
Although this investigation certainly needs refinement, we already appreciate a signature of the different curvature evolution that furnishes a clearcut prediction for future observations (see \cite{morphon:obs} for details).

\medskip\medskip\medskip

\begin{acknowledgments}
{\sl Thanks go to my collaborators with whome I share some of the presented results. 
It is a pleasure to thank the organizers of the workshop for inviting this contribution.
This work is supported by ``F\'ed\'eration de Physique Andr\'e--Marie Amp\`ere'' of Universit\'e Lyon 1 and \'Ecole Normale Sup\'erieure de Lyon. }
\end{acknowledgments}
\bigskip\bigskip

\end{document}